\documentclass[conference]{IEEEtran}
\IEEEoverridecommandlockouts
\usepackage{cite}
\usepackage{amsmath,amssymb,amsfonts}
\usepackage{algorithmic}
\usepackage{graphicx}
\usepackage{textcomp}
\usepackage[hang,small,bf]{caption}
\usepackage[subrefformat=parens]{subcaption}
\usepackage{multirow}
\usepackage{xcolor}
\def\BibTeX{{\rm B\kern-.05em{\sc i\kern-.025em b}\kern-.08em
    T\kern-.1667em\lower.7ex\hbox{E}\kern-.125emX}}
\begin{document}

\title{Attack Tree Analysis for Adversarial Evasion Attacks}

\author{\IEEEauthorblockN{Yuki Yamaguchi}
\IEEEauthorblockA{
\textit{Japan Advanced Institute of Science and Technology}\\
1-1, Asahidai, Nomi, 923-1292, Ishikawa, Japan \\
s2110177@jaist.ac.jp}
\and
\IEEEauthorblockN{Toshiaki Aoki}
\IEEEauthorblockA{
\textit{Japan Advanced Institute of Science and Technology}\\
1-1, Asahidai, Nomi, 923-1292, Ishikawa, Japan \\
toshiaki@jaist.ac.jp}
}

\maketitle
\begin{abstract}
Recently, the evolution of deep learning has promoted the application of machine learning (ML) to various systems.
However, there are ML systems, such as autonomous vehicles, that cause critical damage when they misclassify.
Conversely, there are ML-specific attacks called adversarial attacks based on the characteristics of ML systems.
For example, one type of adversarial attack is an evasion attack, which uses minute perturbations called "adversarial examples" to intentionally misclassify classifiers.
Therefore, it is necessary to analyze the risk of ML-specific attacks in introducing ML base systems.
In this study, we propose a quantitative evaluation method for analyzing the risk of evasion attacks using attack trees.
The proposed method consists of the extension of the conventional attack tree to analyze evasion attacks and the systematic construction method of the extension.
In the extension of the conventional attack tree, we introduce ML and conventional attack nodes to represent various characteristics of evasion attacks.
In the systematic construction process, we propose a procedure to construct the attack tree.
The procedure consists of three steps: (1) organizing information about attack methods in the literature to a matrix, (2) identifying evasion attack scenarios from methods in the matrix, and (3) constructing the attack tree from the identified scenarios using a pattern.
Finally, we conducted experiments on three ML image recognition systems to demonstrate the versatility and effectiveness of our proposed method.

\end{abstract}


\section{Introduction}
\begin{figure}[hbtp]
    \centering
    \begin{minipage}[b]{80mm}
        \includegraphics[width=80mm]{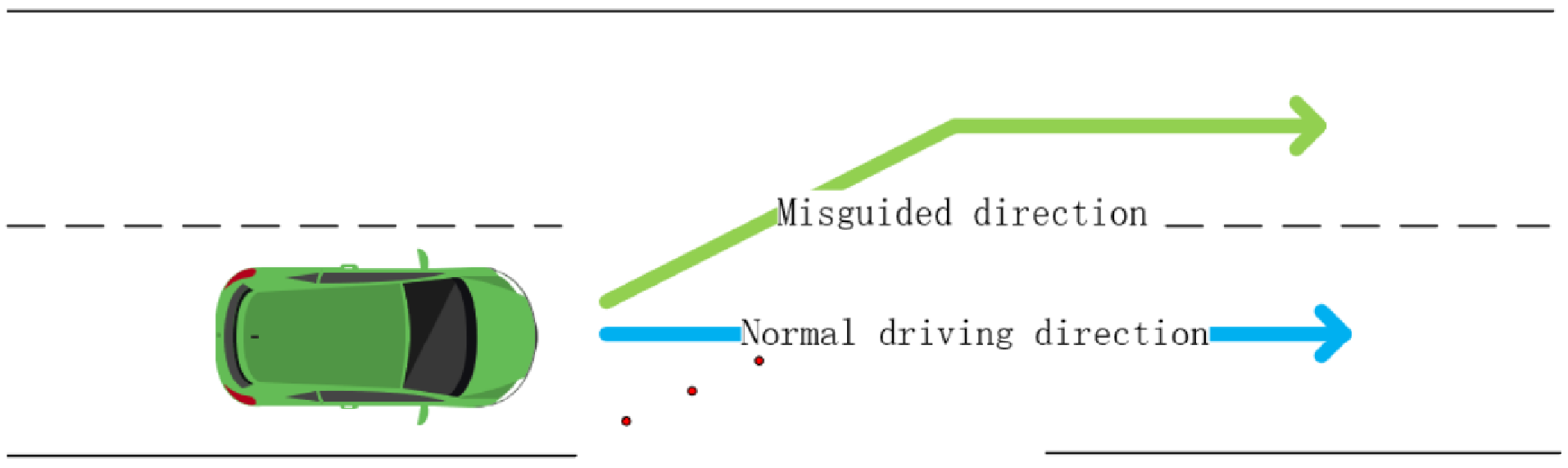}
    \end{minipage}\\
    \vspace{1pc}
    \begin{minipage}[b]{80mm}
        \includegraphics[width=80mm]{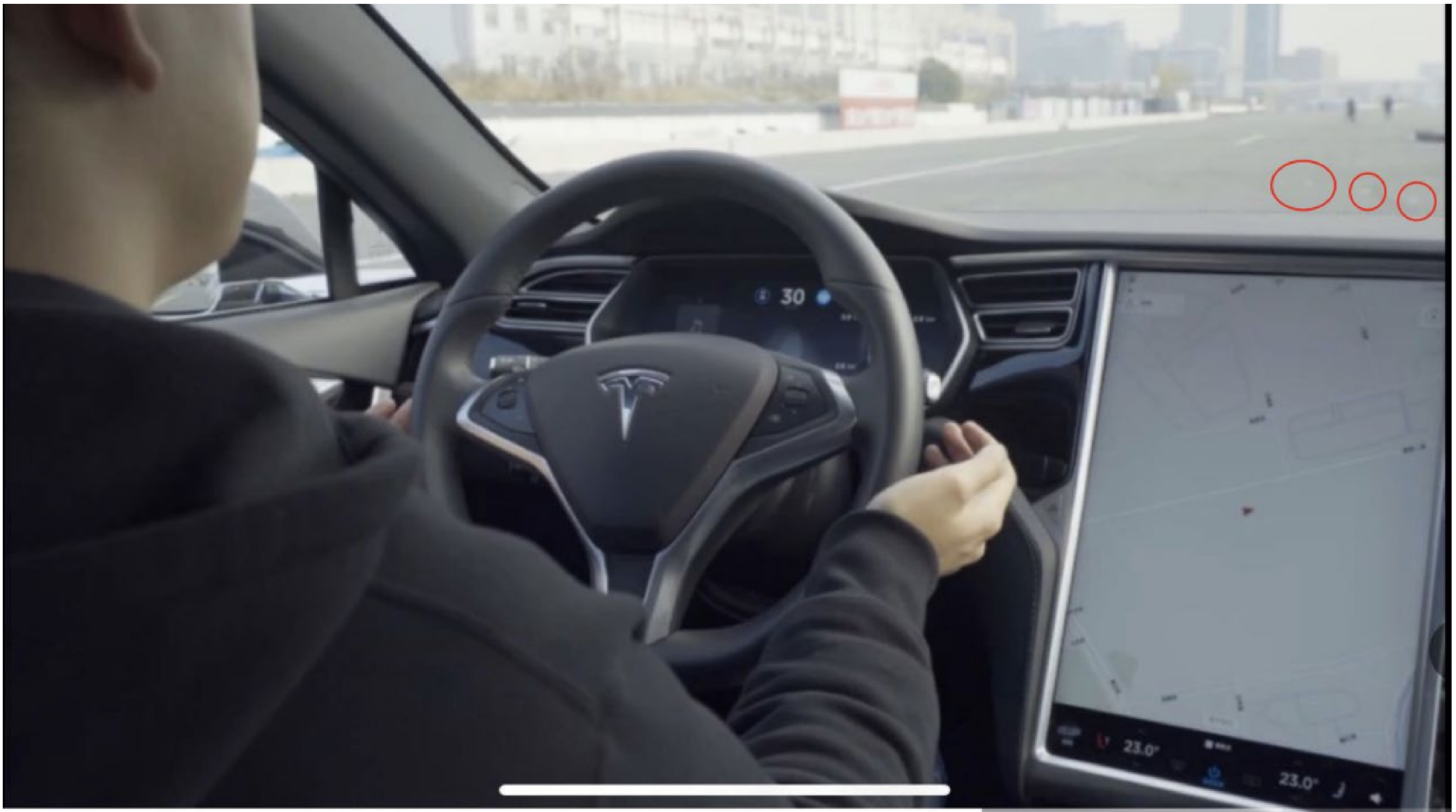}
    \end{minipage}
    \caption{Evasion attack using physical adversarial examples made the Tesla autopilot function change to an opposing lane in the experiment\cite{tencent2019experimental}.}
    \label{fig:EAExperiment}
\end{figure}

Because of advancements in machine learning (ML), various systems now include ML components.
Several ML systems are safety-critical such as autonomous driving.
Conversely, some ML-specific vulnerabilities result from the characteristics of ML.
For example, adversarial example vulnerability\cite{szegedy2013intriguing} causes the misclassification of an ML classifier using a minute perturbation generated intentionally.
Evasion attacks are adversarial attacks that cause a classifier to misclassify using an adversarial example. Tencent Security Lab conducted an evasion attack experiment in a Tesla car autopilot that causes lane change to opposing lane using physical adversarial examples such as three small circles on the road\cite{tencent2019experimental} (Fig. \ref{fig:EAExperiment}).
Therefore, it is necessary to analyze the risk of evasion attacks in introducing ML base systems.
In this study, we propose a quantitative evaluation method for analyzing the risk of evasion attacks on ML systems.

Attack tree analysis\cite{Schneier1999AttackTree} is a typical method for analyzing attacks on systems.
It represents the attack hierarchically using a tree structure.
An attack tree has various methods for computing security metrics such as the probability of a successful attack\cite{bagnato2012attribute, buldas2020attribute} or the minimum cost of an attack\cite{Schneier1999AttackTree}.
Let us consider analyzing evasion attacks using conventional attack (CA) trees.
Analysts set leaf nodes to determine the probability that the attacks succeed and compute the attributes of the parent nodes bottom up in computing the probability.
Conversely, an adversarial example method (AEM) used for evasion attacks computes the error rate of the classifier experimentally.
Thus, attack trees need to handle those values separately.
An attack library such as the adversarial robustness toolbox (ART)\cite{nicolae2018ART} has organized some of these AEMs, but many experimentally coded methods can be found on GitHub.
Furthermore, the evasion attack difficulty depends on the characteristic of the AEM used.
In addition, calculating the minimum cost as the query that requires evasion attacks from AEMs is useful for determining mitigation.
It is impossible to handle the characteristics above using CA trees.
Therefore, we propose extensions of attack trees to address this issue.
Our method consists of the nodes of attack trees to separate ML and conventional components and the calculation of the attack probability (AP) and the minimum query for an evasion attack.

The effectiveness of attack tree analysis depends on the knowledge of analysts.
Therefore, the analyst shall be familiar with the security domain.
Furthermore, the analysis of evasion attacks requires both security- and ML-domain knowledge because of ML-specific vulnerabilities such as adversarial examples.
However, a survey of industrial practitioners found that both ML engineers and incident responders are unequipped to secure ML systems against adversarial attacks\cite{kumar2020AMLIndustryPerspective}.
Therefore, there is a need for a method for systematically building an attack tree of evasion attacks to support analysts.
In this study, we provide a systematic construction procedure of an attack tree to increase the effectiveness of attack tree analysis.

Our method consists of attack tree extensions for evasion attacks and a systematic construction method for attack trees.
We describe the features of our methods.
Our extensions of attack trees separate evasion attacks into ML and conventional elements.
They allow the handling of the characteristics of evasion attacks unique to ML such as error rates and method attributes.
Thus, our attack trees can evaluate evasion attacks more appropriately than conventional ones.
Furthermore, attack trees for ML attacks require the capabilities of security and ML domains.
Therefore, analysts have more burdens to construct attack trees for evasion attacks than CAs.
Our systematic construction method helps analysts create an attack tree and analyze an evasion attack.

We conducted experiments to evaluate the versatility and effectiveness of our methods.
We analyzed three evasion attacks on different image classification systems using our methods.
The results suggest that our methods have versatility for evasion attacks in image classification systems.
Furthermore, the experiments indicate that our methods are effective in reducing the risk of evasion attacks.
We could consider the trade-off of each mitigation through calculated APs.
Minimum queries for evasion attacks in black-box settings determine the parameter of query limit mitigation.

This paper is organized as follows:
In section II, we show related works.
In section III, we describe the ideas of our method and systematic construction method of attack trees.
In section IV, we detail the attack tree extension for evasion attacks and describe the computation of the probability of evasion attacks.
In section V, we describe the experiment for evaluating our method.
In section VI, we discuss the advantages and limitations of our method.
In section VII, we conclude this study and present future works.

\section{Related Work}
In this section, we describe the vulnerability of adversarial examples and evasion attacks using adversarial examples.
Furthermore, we refer to the security literature or tools for such ML attacks and introduce a CA analysis method, i.e., attack trees.

Adversarial examples are minute noises that humans cannot recognize but classifiers misclassify by.
Researchers have proposed many types of methods for generating them.
The fast gradient sign method (FGSM)\cite{goodfellow2014FGSM} can generate adversarial examples fast.
The robust physical perturbations (RP2)\cite{Eykholt2018RP2} create physical adversarial examples such as stickers.
These methods generate noises using the gradient of the classifier.
Thus, generating adversarial examples in these white-box setting attacks requires full information about the classifier.
Conversely, there are methods for creating adversarial examples without the classifier gradient\cite{brendel2017BoundaryaAttack}\cite{ guo2019SimBA}\cite{chen2020hopskipjumpattack}.
These methods create adversarial examples from the tuple of the input and output of the classifier.
There is a comprehensive survey for the aforementioned AEMs\cite{machado2021AMLSurvey}.
It uses the taxonomy of the attributes of the method such as perturbation visibility and computation to generate adversarial examples.
Currently, no consensus knowledge such as common vulnerabilities and exposures (CVE)\cite{CVE} in conventional security base organizes information about adversarial examples\cite{kumar2020AMLIndustryPerspective}.
Therefore, we extract information about attributes from the literature and organize them into a matrix in this study.

Each AEM hypothesizes the model accessibility of attackers.
The white-box setting\cite{machado2021AMLSurvey} is a strong hypothesis that attackers can completely access the victim model and generate adversarial examples with the model gradient.
The black-box setting\cite{papernot2017practicalBBAttack} is a feasible hypothesis that attackers can only conduct queries to the model and generate adversarial examples with the model input and output.
Furthermore, there is a distinguished black-box attack using the proxy model.
The proxy model learns the training set of a similar distribution to the victim dataset.
The research\cite{papernot2016transferability} suggests that adversarial examples generated using the proxy model also typically misclassify the victim model.
From above, attackers could conduct evasion attacks in a black-box setting easier than in a white-box setting.
Therefore, our method includes the calculation for the minimum queries for evasion attacks in a black-box setting. 

Adversarial attacks exploit ML-specific vulnerabilities such as adversarial examples.
They consist of four attacks.
Evasion attacks misclassify the victim model with intensity using adversarial examples.
Data poisoning causes the victim model to learn the poisoned model using intended training data\cite{cina2022DPSurvey}.
Model extraction steals information about the model from the pair of input and output\cite{tramer2016stealing}.
Data extraction steals information about the training data in the same way as model extraction\cite{shokri2017membership}.
In this research, we mainly analyze the risk of evasion attacks.

There is little ML security research on evasion attacks for now.
The ATLAS framework\cite{ATLAS} describes adversarial attack procedures using the method of tactics, techniques, and procedures (TTPs).
It does not contain information about the methods used by attackers.
Therefore, ATLAS provides no identification of the methods used for attacks.
The adversarial robustness toolbox (ART)\cite{nicolae2018ART} organizes various methods for adversarial attacks as a practical library for security experiments.
It contains many unique methods of adversarial examples.
We experimented with the ART library in this study.

\section{Approach}
\subsection{Attack trees for evasion attacks}
Evasion attacks consist of generating adversarial examples and injecting them into the victim model.
Thus, we need to consider methods for generating adversarial examples in the analysis of evasion attacks.
For example, attackers use physical attribute methods to generate physical adversarial examples such as stickers to evade a model.
Conversely, attackers use digital attribute methods when they input adversarial examples directly.
The frequency of an evasion attack with a physical or digital adversarial example could change depending on the victim model and the analyzed evasion attack.
Furthermore, adversarial examples have a model error rate and frequency, as well as implementation difficulty.
However, existing attack trees are inappropriate to represent these characteristics of evasion attacks.
Therefore, we propose an extension of attack trees for evasion attacks (AT4EA).
Our attack trees include a node to represent domain knowledge of method attributes.
AT4EA also extends nodes to handle the error rate of the method and its implementation difficulty.
By the extension above, we handle evasion attacks by separating the attacks into the ML and conventional domains.
In addition, we propose the computation of the AP and minimum queries for evasion attacks using AT4EA.

We described AT4EA in detail in section IV.

\subsection{Systematic Attack Tree Construction}
The effectiveness of an attack tree depends on the analyst's knowledge.
For example, knowledgeable analysts can identify various attack scenarios to achieve the goal and construct an attack tree that represents these scenarios.
It is difficult for conventional security analysts to construct effective attack trees in evasion attack analysis because it requires knowledge of security and ML. In this study, we propose systematic attack tree construction to develop effective attack trees for evasion attacks.

\begin{figure}[hbtp]
    \centering
    \includegraphics[width=0.9\linewidth]{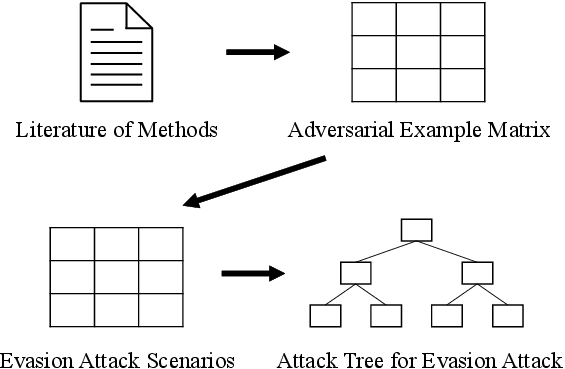}
    \caption{ Overview of our methods}
    \label{fig:Approach}
\end{figure}

We show an overview of our systematic construction procedure in Fig. \ref{fig:Approach}.

First, organizing information about AEMs from each literature into an AEM matrix (AEMMatrix).
There is no comprehensive knowledge base of AEMs such as conventional CVE.
Therefore, we need to extract their information from the literature.
Second, discovering evasion attack scenarios (EAS) covering all methods in AEMMatrix.
Analysts can describe EAS with tactics, techniques, and procedures like ATLAS.
However, ATLAS shows attack scenarios without information about AEMs.
Therefore, it is impossible to describe the methods used for an ATLAS scenario.
For example, attackers can use physical adversarial examples and cannot use digital ones in an EAS such as road sign misclassification with physical stickers.
We introduce the EAS to represent such relations between scenarios and methods.
EAS includes TTP-based scenarios and the available methods for them.
Third, constructing AT4EA from each EAS.
The analysts can perform attack tree analysis by constructing AT4EA from identified EAS.
We propose a pattern for translating from EAS to AT4EA.
Our pattern is based on the idea that evasion attacks consist of adversarial example generation and insertion.

The details of our construction method are described in section V.

\section{Attack Trees for Evasion Attacks}
\subsection{Nodes}
\subsubsection{Definitions}
We describe the purpose of nodes.
AT4EA separates an evasion attack into ML and conventional elements to handle the characteristics of ML.
Thus, we extend new nodes for ML components such as AEMs and AEM attributes.
Furthermore, our extension has an EAS node for constructing AT4EA systematically, as described in section V.
Now, we introduce nodes for representing evasion attacks.
These include root, adversarial example attribute (AEA), scenario, AEM list (AEML), AEM, CA list (CAL), and CA nodes.
We show the general form of these nodes below.

\begin{figure}[hbtp]
    \begin{tabular}{cc}
    \begin{minipage}[b]{0.35\linewidth}
        \centering
        \includegraphics[width=0.8\linewidth]{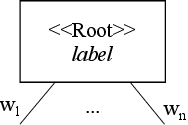}
        \subcaption{Normal form}\label{fig:RootNodeDefinition.normal}
    \end{minipage}
    \begin{minipage}[b]{0.65\linewidth}
        \centering
        \includegraphics[width=0.8\linewidth]{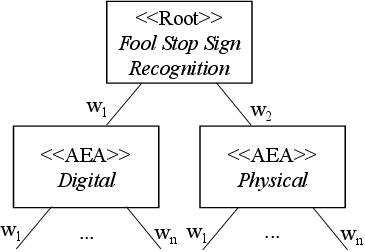}
        \subcaption{Example}\label{fig:RootNodeDefinition.example}
    \end{minipage}
    \end{tabular}
    \caption{(a) shows the normal form of the root node. (b) exemplifies the root node.}
\end{figure}

\paragraph{Root node}
Fig. \ref{fig:RootNodeDefinition.normal} show the normal form of the root node.
It is the root of the attack tree.
Its label means the evasion attack objective, the same as the root of CA trees.
Let us consider an evasion attack against a road sign recognition system.
We analyze the attack that misclassifies the stop sign using AT4EA.
Fig. \ref{fig:RootNodeDefinition.example} shows the root part of the tree representing the evasion attack.
There are two cases where an adversarial example is used in a digital or physical domain.
Therefore, we can represent the evasion attack using the root node by breaking it down into digital or physical representations using the AEA node.
To add an adversarial example, the attack in the physical domain requires adding stickers on the road sign.
Conversely, the attack in the digital representation needs real-time adversarial example generation and insertion between the vehicle cameras and the classifier.
Thus, we could consider that the attack in the physical domain occurs more frequently than in the digital domain in this case.
The weight of edges between the root node and its children represents the frequency of evasion attacks to represent domain knowledge.

\begin{figure}[hbtp]
    \begin{tabular}{cc}
    \begin{minipage}[b]{0.35\linewidth}
        \centering
        \includegraphics[width=0.8\linewidth]{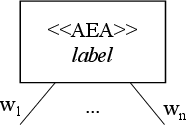}
        \subcaption{Normal form}\label{fig:AEANodeDefinition.normal}
    \end{minipage}
    \begin{minipage}[b]{0.65\linewidth}
        \centering
        \includegraphics[width=0.8\linewidth]{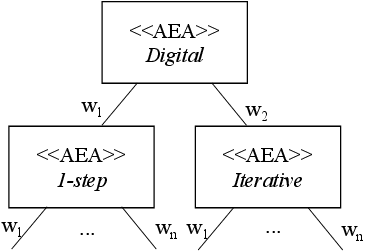}
        \subcaption{Example}\label{fig:AEANodeDefinition.example}
    \end{minipage}
    \end{tabular}
    \caption{(a) shows the normal form of the AEA node. (b) exemplifies its use.}
\end{figure}

\paragraph{AEA node}
Fig. \ref{fig:AEANodeDefinition.normal} shows the normal form of the AEA node.
It represents the attribute of AEMs.
The AEA node branches evasion attacks based on AEM attributes.
For example, Fig. \ref{fig:AEANodeDefinition.example} shows that the AEA node  divides the digital AEA node into 1-step and iterative attributes.
In the subtree of these AEA nodes, attacks require an AEM with attributes.
The subtree of the 1-step AEA node in Fig. \ref{fig:AEANodeDefinition.example} has attacks with 1-step attribute methods such as FGSM but not iterative ones such as CW attack\cite{carlini2017CWAttack}.
The weight of AEA node edges represents the frequency of attacks, as in the root node.
Using the AEA node, we can consider an EAS for each method attribute.

\begin{figure}[hbtp]
    \begin{tabular}{cc}
    \begin{minipage}[b]{0.5\linewidth}
        \centering
        \includegraphics[width=0.8\linewidth]{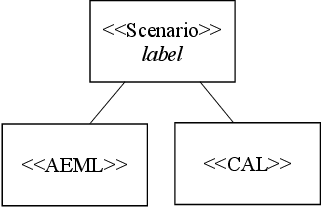}
        \subcaption{Scenario}\label{fig:ScenarioNodeDefinition.scenario}
    \end{minipage}
    \begin{minipage}[b]{0.25\linewidth}
        \centering
        \includegraphics[width=0.8\linewidth]{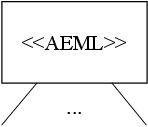}
        \subcaption{AEML}\label{fig:ScenarioNodeDefinition.aeml}
    \end{minipage}
    \begin{minipage}[b]{0.25\linewidth}
        \centering
        \includegraphics[width=0.8\linewidth]{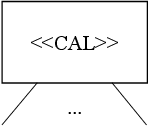}
        \subcaption{CAL}\label{fig:ScenarioNodeDefinition.cal}
    \end{minipage}
    \end{tabular}
    \caption{Normal forms of scenario, AEML, and CAL nodes: (a) scenario, (b) AEML, and (c) CAL.}
\end{figure}

\paragraph{Scenario, AEML, and CAL nodes}
Fig. \ref{fig:ScenarioNodeDefinition.scenario} shows the normal form of the scenario node.
It represents an EAS based on TTPs and AEMs, which can be used for an evasion attack.
An EAS is a component for systematic AT4EA construction and is described in detail in section V.
The scenario node has the AEML node shown in Fig. \ref{fig:ScenarioNodeDefinition.aeml} and the CAL node shown in Fig. \ref{fig:ScenarioNodeDefinition.cal} as its children.
The AEML node has AEMs that can be used for the scenario as AEM nodes.
It succeeds when at least one of its children fools the target classifier.
Conversely, the CAL node has CAs required for the scenario as CA nodes.
It succeeds when all its children complete their goals.
The scenario node completes its scenario when the AEML and CAL nodes succeed.

\begin{figure}[hbtp]
    \begin{tabular}{cc}
    \begin{minipage}[b]{0.31\linewidth}
        \centering
        \includegraphics[width=0.8\linewidth]{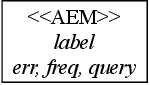}
        \subcaption{Normal form}\label{fig:AEMNodeDefinition.normal}
    \end{minipage}
    \begin{minipage}[b]{0.69\linewidth}
        \centering
        \includegraphics[width=0.8\linewidth]{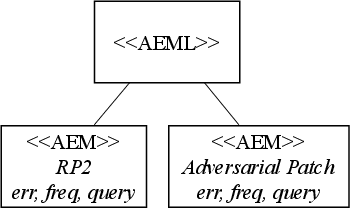}
        \subcaption{Example}\label{fig:AEMNodeDefinition.example}
    \end{minipage}
    \end{tabular}
    \caption{(a) shows the normal form of the AEM node. (b) exemplifies its use.}
\end{figure}

\paragraph{AEM node}
Fig. \ref{fig:AEMNodeDefinition.normal} shows the normal form of the AEM node.
It represents generation methods for adversarial examples.
The AEM node succeeds if the adversarial example it generates fools the target.
Let us consider the case that the AEML node in Fig. \ref{fig:AEMNodeDefinition.example} belongs to a subtree of a physical scenario.
The AEML node has two AEM nodes, namely RP2 and adversarial patch\cite{brown2017AdvPatch}.
The AEML node succeeds if the victim classifier obtains an error using any adversarial example generated by the RP2 or adversarial patch node.

\begin{figure}[hbtp]
    \begin{tabular}{cc}
    \begin{minipage}[b]{0.32\linewidth}
        \centering
        \includegraphics[width=0.8\linewidth]{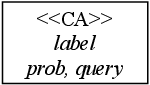}
        \subcaption{Normal form}\label{fig:CANodeDefinition.normal}
    \end{minipage}
    \begin{minipage}[b]{0.68\linewidth}
        \centering
        \includegraphics[width=0.8\linewidth]{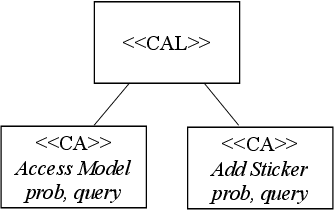}
        \subcaption{Example}\label{fig:CANodeDefinition.example}
    \end{minipage}
    \end{tabular}
    \caption{(a) shows the normal form of the CA node. (b) exemplifies its use.}
\end{figure}

\paragraph{CA node}
Fig. \ref{fig:CANodeDefinition.normal} shows the normal form of the CA node.
It represents CAs for inputting an adversarial example for an evasion attack.
Let us consider the sticker attack scenario that fools the road sign recognition system by placing adversarial stickers on the stop sign.
It requires the conventional attacks classifier access and stickers, as shown in Fig. \ref{fig:CANodeDefinition.example}.
The CAL node succeeds if all attacks are complete.
We can branch CA trees under the CA node to model the evasion attack in detail.

\subsubsection{AT4EA Example}
\begin{figure}[hbtp]
    \centering
    \includegraphics[width=0.95\linewidth]{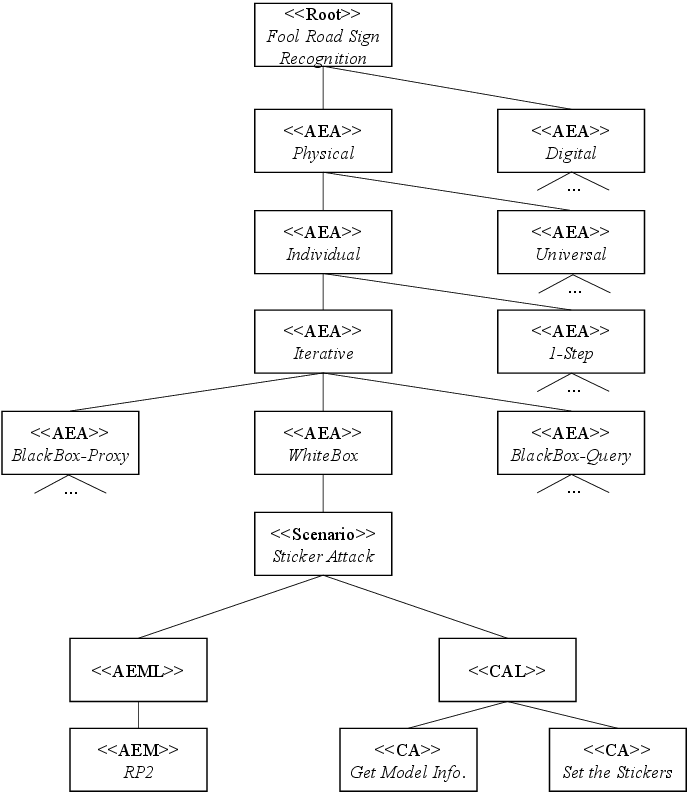}
    \caption{Example of AT4EA: The sub-tree of AT4EA for the road sign recognition system in section VI.} \label{fig:NodesExample}
\end{figure}

We use the sub-tree of AT4EA for the road sign recognition system in the section VI experiment to demonstrate our method in Fig. \ref{fig:NodesExample}.
The physical and digital AEA nodes belong to the root node as its children.
These are the AEM attributes.
The below of the physical AEA node means the same as the method attributes.
The scenario node under the white-box AEA represents an EAS.
It has an RP2 AEM node as the method attackers can use in this scenario because the attributes of the AEM nodes are equal to the physical, individual, iterative, and white-box AEA nodes of the upper of the scenario node.
It also includes the two CA nodes as the attack steps to complete the scenario.

AT4EA represents the following.
The sticker attack scenario has the restriction of upper AEA nodes.
For example, this EAS suggests that the attack proceeds using physical adversarial examples, which are individually generated not in real time and in the white-box setting.
The attack steps of the sticker attack scenario consist of collecting model information to generate adversarial examples, creating them, and setting them into the road sign as intentional stickers.
In this example, the method for generating adversarial examples is RP2.
These attributes must equal the AEA nodes between the root and scenario nodes.
Our AT4EA analyzes evasion attacks in the above way.

\subsection{Attribute Calculation}
We introduce two attribute calculations to quantitatively analyze evasion attacks by AT4EA.
They consist of the AP and minimum query required for a black-box attack.
The AP indicates the vulnerability of ML systems to evasion attacks.
The minimum query can identify the parameters of query limit mitigations for black-box attacks.
Computing these values using AT4EA helps analysts consider the trade-off of mitigations and identify the query limit mitigation for black-box attacks.
We demonstrate the attribute calculations for each node below.

\subsubsection{Attack Probability}
The AP represents the frequency with which the evasion attack succeeds.
It suggests the vulnerability of ML systems to evasion attacks.
For example, we could consider the trade-off of the mitigation for the evasion attack monitoring the AP of the system.
We show the calculation of the AP by each node below.

\begin{align}
    \label{eq:ap.aem}
    ap(aem) = err \times freq
\end{align}
\begin{align}
    \label{eq:ap.aeml}
    ap(aeml) = max \{ap(m) | m \in C\}
\end{align}
\begin{align}
    \label{eq:ap.ca}
    ap(ca) = prob
\end{align}
\begin{align}
    \label{eq:ap.cal}
    ap(cal) = \prod_{m \in C} ap(m)
\end{align}
\begin{align}
    \label{eq:ap.scenario}
    ap(sc) = ap(aeml) \times ap(cal)
\end{align}
\begin{align}
    \label{eq:ap.aea}
    ap(n) = \sum^{l}_{i=1} w_i \times ap(c_i)
\end{align}

%% AEM
\paragraph{AEM node}
We show the AP of the AEM node $ap(aem)$ calculated from its $err$ and $freq$ values in Eq \ref{eq:ap.aem}.
$err$ denotes the experimental error rate against the victim model caused by the AEM.
$freq$ denotes a value between 0 and 1 representing the attack frequency.
This frequency represents the difficulty of the attack using the method.
Let us consider the two methods FGSM and RP2.
We can easily code the FGSM method in various ML development tools using evasion attack libraries such as ART.
Conversely, the RP2 method is only available on GitHub as an experimental code for its literature, and we need to build the attack program from it.
Thus, the FGSM $freq$ value should be greater than the RP2 $freq$ value in this case.
We define the AP of the AEM node $ap(aem)$ as the multiplication of $err$ and $freq$.

%% AEML
\paragraph{AEML node}
Next, we consider the AP of the AEML node $ap(aeml)$ in Eq \ref{eq:ap.aeml}.
The AEML node succeeds when one of its children fools the victim model.
In addition, we hypothesize that attackers choose the most successful method for an evasion attack.
Therefore, we define the AP of AEML $ap(aeml)$ as the maximum $ap(aem)$ of its children $C$.

%% CA
\paragraph{CA node}
We consider the AP of the CA node $ap(ca)$ in Eq \ref{eq:ap.ca}.
In conventional security attacks, we assume that the AP assigned to leaf nodes of attack trees can be calculated from past data such as annual crime reports.
The CA node represents the attack in the context of conventional security, not the ML one.
Therefore, we assign the conventional probability to the $prob$ value of CA nodes or calculate it using CA trees.

%% CAL
\paragraph{CAL node}
Next, we consider the AP of the CAL node $ap(cal)$ in Eq \ref{eq:ap.cal}
The CAL node succeeds when all its children CA nodes achieve their goals.
Thus, we defined the calculation of the AP of the CAL node $ap(cal)$ as the simultaneous probability of $ap(ca)$ of its children $C$.

%% Scenario
\paragraph{Scenario node}
We consider the AP of the scenario node $ap(sc)$ in Eq \ref{eq:ap.scenario}.
The scenario node has only two children, AEML and CAL nodes.
The scenario node succeeds when both nodes are successful.
Therefore, we defined the calculation of the AP of the scenario $ap(sc)$ as the multiplication of $ap(aeml)$ and $ap(cal)$.

%% Root and AEA
\paragraph{AEA and root nodes}
Finally, we consider the AP of the AEA and root nodes $ap(n)$ in Eq \ref{eq:ap.aea}.
These nodes succeed when their children achieve their goals.
Furthermore, they have weight values $w_i$ on the edges of their children.
They represent the frequency that means the domain knowledge from the attributes of the AEMs.
Thus, we defined the calculation of the AEA and root nodes $ap(n)$ by their $l$ children's probability $ap(c_i)$ and frequency $w_i$ for vulnerability evaluation of the evasion attack against the system.

\subsubsection{Minimum query required for black-box attack}
In black-box settings, attackers can access the model using only a query.
Thus, they generate adversarial examples using queries.
Furthermore, they can collect information about the model using the query to construct a proxy model.
Calculating the minimum query for evasion attacks is effective in identifying the mitigation parameter that restricts them.

In the calculation of the minimum query, we hypothesize using the black-box setting because white-box methods generate adversarial examples without queries.
Therefore, the calculation is applied to AT4EA without subtrees below the AEA nodes labeled white-box.

\begin{align}
    \label{eq:mq.aem}
    mq(aem) = query
\end{align}
\begin{align}
    \label{eq:mq.aeml}
    mq(aeml) = min \{mq(aem) | aem \in C\}
\end{align}
\begin{align}
    \label{eq:mq.ca}
    mq(ca) = query
\end{align}
\begin{align}
    \label{eq:mq.cal}
    mq(cal) = \sum_{ca \in C} mq(ca)
\end{align}
\begin{align}
    \label{eq:mq.scenario}
    mq(sc) = mq(aeml) + mq(cal)
\end{align}
\begin{align}
    \label{eq:mq.aea}
    mq(n) = min \{mq(m) | m \in C\}
\end{align}

%% AEM
\paragraph{AEM node}
We show the minimum query of the AEM node $mq(aem)$ in Eq \ref{eq:mq.aem}.
The parameter $query$ represents queries for generating adversarial examples by the method of the AEM node.
Let us consider evasion attacks in black-box settings.
An attacker can only input into the model and receive an output.
In this situation, the attacker can use black-box AEMs with queries.
These black-box AEMs, such as Boundary Attack, create adversarial examples by the model output iteratively, whereas white-box methods, such as FGSM, create them by the model gradient.
We define the minimum query of the AEM node $mq(aem)$ as the $query$ value for generating adversarial examples and assign it to the AEM node.

%% AEML
\paragraph{AEML node}
Next, we consider the minimum query of the AEML node $mq(aeml)$ in Eq \ref{eq:mq.aeml}.
The AEML node succeeds when the method $aem$ of its children $C$ fools the victim model.
Furthermore, we hypothesize that attackers prefer to use methods that require a minimum query for attacks in black-box settings. 
Thus, we define the minimum query of the AEML node $mq(aeml)$ as the least minimum query of its children $mq(aem)$.

%% CA
\paragraph{CA node}
We consider the minimum query of the CA node $mq(ca)$ in Eq \ref{eq:mq.ca}.
Many CAs achieve their goals without queries.
However, some attacks apply queries to the victim model such as attacks using a proxy model.
For example, the attacker may use a query to discover the output space of the victim model or to extract its information.
Thus, we assign such required queries to the $query$ of the CA node.
Then, we define the minimum query of the CA node $mq(ca)$ as the query for the CA.

%% CAL
\paragraph{CAL}
Next, we consider the minimum query of the CAL node $mq(cal)$ in Eq \ref{eq:mq.cal}.
The CAL node succeeds when all its children $ca$ achieve their goals.
Therefore, we define the minimum query of the CAL node $mq(cal)$ as the sum of the minimum query of the CA nodes $mq(ca)$ of its children $C$.

%% Scenario
\paragraph{Scenario node}
We show the minimum query of the scenario node $mq(sc)$ in Eq \ref{eq:mq.scenario}.
The scenario node has just two children, AEML and CAL nodes.
The scenario node succeeds when its two children achieve their goals.
Thus, we define the minimum query of the scenario node $mc(sc)$ as the sum of the minimum query of the AEML node $mc(aeml)$ and the CAL node $mq(CAL)$.

%% AEA and Root
\paragraph{AEA and root nodes}
Finally, we consider the minimum query of the AEA and root nodes $mq(n)$ in Eq \ref{eq:mq.aea}.
They succeed when at least one child $m$ of their children $C$ succeeds.
In the hypothesis that an attacker prefers fewer queries for evasion attacks, the minimum query of the AEA and root nodes is calculated using the least minimum query of their children $mq(m)$.
Therefore, we define the minimum query of the AEA and root nodes $mq(n)$ as Eq \ref{eq:mq.aea}.

\subsubsection{Attribute Calculation Example}
\begin{figure*}[hbtp]
    \centering
    \includegraphics[width=0.95\linewidth]{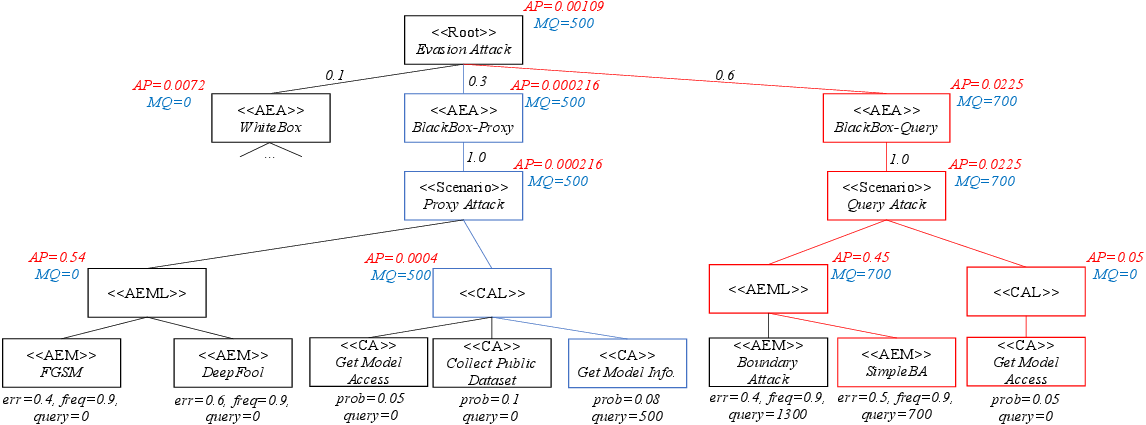}
    \caption{Example of attribute calculation. The red numbers are the AP. The blue numbers are the minimum queries for evasion attacks.} \label{fig:CalcExample}
\end{figure*}

We show an example of attribute calculation for AT4EA in Fig. \ref{fig:CalcExample}.
We describe each calculation below.

The red colors mean the AP calculation.
The AP of AEM nodes means the multiplication of its error rate $err$ and frequency $freq$.
Then, the AP of the AEML nodes is the maximum of them.
The AP of CAL equals the multiplication of probabilities $prob$ of its children.
The scenario AP multiplies the AEML and CAL APs.
In this example, each AEA node has only one scenario; thus, its AP equals that of its child.
Finally, the AP of the root node equals the sum of the multiplication of edge weights and its children.
Furthermore, we color the nodes that most contribute to the evasion attack red.
This AT4EA suggests that the query attack with the SimpleBA\cite{guo2019SimBA} method is the most successful scenario.
The AP of the root node measures the vulnerability to evasion attacks.
Calculating it for the system used for some mitigations allows us to evaluate its effectiveness and weigh the trade-offs.
The blue color indicates the calculation of the minimum queries for the evasion attack (MQ).
The minimum query of AEML nodes chooses the minimum $query$ of their children.
The minimum query of CAL nodes equals the sum of $query$ of their children.
Then, the scenario minimum query is the sum of the AEML and CAL nodes.
In the calculation of the minimum query, we only consider the black-box settings.
Therefore, the calculation of the minimum query does not handle the subtree of White-Box AEA nodes.
Finally, the minimum query of the root and AEA nodes chooses the least minimum query of their children without the white-box subtree.
Furthermore, the nodes that are minimum queries for the evasion attack are indicated in blue.
This minimum query of the root node suggests that the proxy attack scenario requires the least queries for the evasion attack.
Calculating the minimum query helps analysts identify the query limit mitigation parameter for the black-box hypothesis.

\section{AT4EA Construction}
\subsection{Adversarial Example Matrix}
We organize the AEMs for the evasion attacks to construct the AT4EA systematically.
However, there is no practical knowledge base of adversarial examples such as conventional CVE.
Therefore, we introduce the adversarial example matrix to organize the attributes of AEMs from their literature.

We show the normal form of the matrix in the first row in Table \ref{tab:AEMatrix}.
Its columns have the method name and the values of the attribute.
Each row in the matrix represents each AEM.
\begin{table*}[hbtp]
    \centering
    \caption{Row 1: Normal form of adversarial example matrix. Row 3: The example of FGSM\cite{goodfellow2014FGSM} method in the matrix.}
    \label{tab:AEMatrix}
    \begin{tabular}{|l|c|c|c|c|}
    \hline
    Attack & Perturbation Visibility & Perturbation Scope & Attack Computation & Attacker's Knowledge \\ \hline \hline
    name & $att_1$ & $att_2$ & $att_3$ & $att_4$ \\ \hline
    ... & ... & ... & ... & ... \\ \hline
    FGSM\cite{goodfellow2014FGSM} & Digital & Individual & 1-Step & Full \\ \hline
    \end{tabular}
\end{table*}

Let us consider the example of the matrix in the third row in Table \ref{tab:AEMatrix}.
The row represents the FGSM method.
It generates adversarial examples in the digital domain.
They are created from each input of the victim model.
It also calculates them without iteration.
It requires information about the internal operation of the model to generate adversarial examples.
Therefore, it hypothesizes the white-box setting or develops proxy models.
The matrix organizes information about the methods, along with their attributes.

\subsection{Evasion Attack Scenarios}
We show the EAS from the matrix using the TTP style.
Let us consider an evasion attack with several AEMs.
Each method attribute distinguishes different evasion attack procedures.
For example, the process to input adversarial examples into the model differs according to the attribute type such as digital or physical attributes.
However, the existing TTP-based framework for describing EASs does not specify which methods can be used.
Therefore, we introduce an EAS with a pair of its processes and available methods.
Our key idea is that attack procedures are related to AEM attributes.
Evasion attacks have a process to satisfy the hypothesis of the method.
Thus, attack scenarios have the same attributes as the AEMs used in them.
A scenario can use other methods with the same attributes because it satisfies the hypothesis of the methods that have them.

We show the normal form of the EAS in the first row in Table \ref{tab:EAS}.
Its columns have the scenario name, AEM attributes, TTP-based procedures, and available methods for attacks.
Each row in the matrix represents one of the EAS.
\begin{table*}[hbtp]
\centering
\caption{Row 1: Normal form of EAS. Row 3: An EAS in the road sign recognition system in the section VI experiment as the example.}\label{tab:EAS}
\scalebox{0.95}{
\begin{tabular}{|l|c|c|c|c|l|l|}
\hline
\multirow{2}{*}{Scenario} & \multicolumn{4}{c|}{Attributes} & \multirow{2}{*}{Conventional Attack} & \multirow{2}{*}{Available Methods} \\ \cline{2-5}
 & Perturbation Visibility & Perturbation Scope & Attack Computation & Attacker's Knowledge & & \\ \hline \hline
\multirow{3}{*}{name}          & \multirow{3}{*}{$att_1$} & \multirow{3}{*}{$att_2$} & \multirow{3}{*}{$att_3$} & \multirow{3}{*}{$att_4$} & $ca_1$ & $aem_1$ \\
& & & & & ... &  ... \\
& & & & & $ca_n$ & $ca_m$ \\ \hline
... & ... & ... & ... & ... & ... & ... \\ \hline
\multirow{2}{*}{Sticker Attack}          & \multirow{2}{*}{Physical} & \multirow{2}{*}{Individual} & \multirow{2}{*}{Iterative} & \multirow{2}{*}{White-box} & Get Model Info. & \multirow{2}{*}{RP2} \\
& & & & & Set the Stickers & \\ \hline
\end{tabular}
}
\end{table*}
Let us consider one of the scenarios of the road sign recognition system in the section VI experiment as an example in the third row in Table \ref{tab:EAS}.
Sticker is the scenario where the attacker intendedly set stickers on the stop sign to misclassify the model of autonomous vehicles.
Its procedure consists of a process to acquire model information to generate an adversarial example and a process to put the example on the sign as stickers.
In this example, the two methods RP2 and adversarial patch are available for the attack.
In addition to binding the methods to the scenario, the attributes limit how the attack is performed.
For example, adversarial stickers exist in the physical world.
The attacker creates an adversarial example from the target road sign.
This scenario set it in advance.
Therefore, it can compute it iteratively.
This scenario hypothesizes the white-box setting.
We handle the EAS as described above.

\subsection{Pattern}
We propose a pattern for translating the AT4EA from the EAS.
Let us consider the evasion attack on classification systems.
We focused on the components of the evasion attack, including adversarial example generation and insertion into the victim model.
Our AT4EA uses AEMs as the generation process.
Furthermore, the tree structure depends on the insertion process.
For example, it requires interception from the camera to the model in the digital domain, whereas the attacker only sets the sticker in the physical domain.
Thus, these scenarios have different structures in our AT4EA.
The difference in the structures results from the process of the attack scenarios.
Therefore, we introduce a pattern for generating AT4EA from attack scenarios.

\begin{figure}[hbtp]
    \centering
    \includegraphics[width=0.85\linewidth]{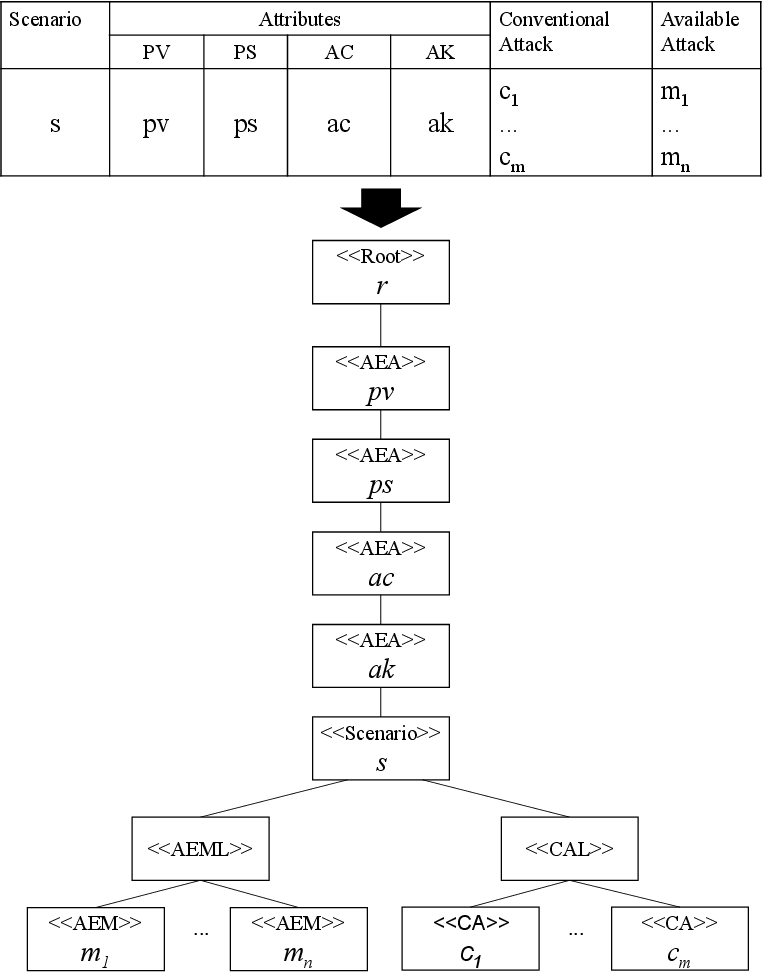}
    \caption{Normal form of the pattern}
    \label{fig:PatternNormalForm}
\end{figure}

We define the pattern in Fig. \ref{fig:PatternNormalForm}.
The evasion attack being analyzed becomes the root node.
The scenario node represents one scenario.
The scenario node has the AEML and CAL nodes as its children.
The attributes of the scenario become the AEA nodes as the root node and itself child in order.
The CA nodes represent the attack process of the scenario and belong to the CAL node.
The AEM nodes represent the available methods for the scenario and belong to the AEML node.

\begin{figure}[hbtp]
    \centering
    \includegraphics[width=0.85\linewidth]{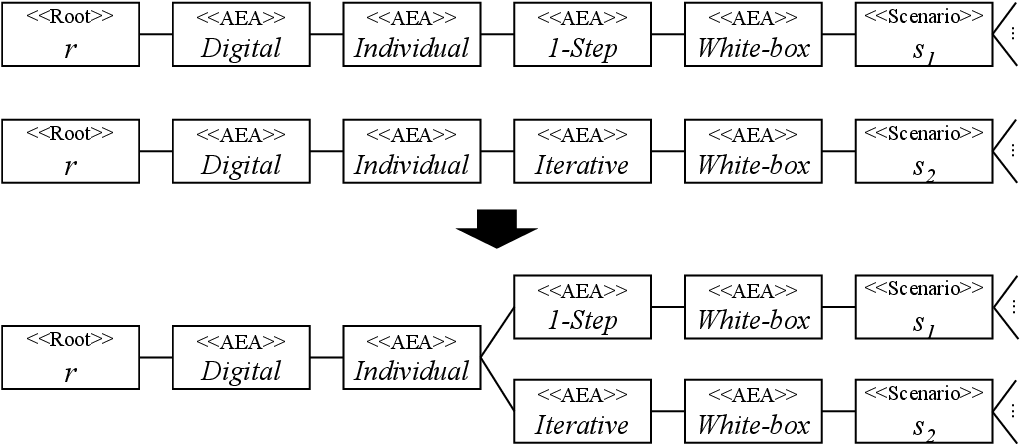}
    \caption{Unify two trees}
    \label{fig:UnifyTrees}
\end{figure}

We show the process that unites the trees of each scenario.
Let us consider the two trees of scenarios A and B in Fig. \ref{fig:UnifyTrees}.
A and B have the same label nodes from the root node to the AEA individual nodes.
The under of the AEA node, the 1-Step and iterative nodes differ.
In this situation of two trees, we add the under of the individual node to the other, as shown in Fig. \ref{fig:UnifyTrees}.
This united tree shows that the attack consists of whether or not it requires generating an adversarial example quickly.
We can construct the AT4EA like this from the trees of each scenario.

\subsection{Construction}
We show the procedure to systematically construct the AT4EA using the matrix, scenarios, and pattern.
First, we extract information about the AEMs from the literature and organize them as a matrix.
Next, we consider the scenarios, including all methods in the matrix.
We acquire the AT4EA from the scenarios translated by the pattern.
The attack procedures as CA nodes in the generated tree are generally abstract.
Therefore, we might need to break them down with CA trees to complete it.
From the above, we systematically obtained the attack tree for the evasion attack.
Finally, we consider the values to assign to the tree from previous works and experiments.
We can analyze the AP and minimum queries for the attack from them.

\section{Experiments}
\subsection{Settings}
We conducted experiments to evaluate the versatility of our method.
In this experiment, our attack trees were constructed systematically to analyze evasion attacks on three systems.
These AT4EA systems evaluated the risk of evasion attacks from the experimental error rates of AEMs.
Three systems consist of road sign recognition with the GTSRB dataset, pneumonia diagnosis with the chest X-ray dataset, and item classification with the fashion MNIST dataset.
We trained the classifiers on these datasets using the Keras framework.
Furthermore, the evasion attacks on these systems are shown below.
The intentional error of stop sign recognition: the attack on human safety during the road sign recognition task.
Wrong diagnosis versus normal diagnosis: a menace to the reliability of an ML provider in the pneumonia diagnosis task.
The intentional misclassification of an item: the mischief to inject poisoned data into the item classification task.
In this experiment, the adversarial example matrix consists of methods with unique attributes.
The EAS are identified, including all methods in the matrix.
Then, our method translates the scenarios to the AT4EA using the pattern.
We train the classifiers to calculate their error rates using adversarial examples in the three different systems.
We calculate the error rates of the classifiers to assess the risk of each system.
The test dataset consists of 50 images.
Each method generates adversarial examples with noise sizes less than the constants.
The codes of the methods consist of 11 programs from the ART library and two programs we wrote.
Furthermore, the methods in the black-box setting assign appropriate query parameters to restrict the size of the adversarial example below a certain value.
The APs of CAs have values between 0.1 and 0.01.
We set the weights of edges to the attacker's knowledge nodes to be white-box: 0.1, black-box (proxy): 0.3, black-box (query): 0.6 or white-box: 0.1, and black-box (proxy): 0.9.
Other weights depend on each system.
The AP of methods packaged in the ART library is 0.9, and that of the others is 0.7.
Under the above conditions, we construct AT4EA using our methods, conduct evasion attacks with 13 AEMs, and analyze them by calculating their attributes from experimental values.

Next, we describe an experiment to evaluate the effectiveness of our method.
We identify four mitigations of the item classification system.
The experiment shows the trade-off of the mitigations by the AP calculated using AT4EA.
We describe the mitigations below.
\begin{itemize}
    \item Adversarial Training (AT)\cite{tramer2017AdversarialTraining}: The defense technique of the ML classifier. We calculate the error rates against the classifier + AT.
    \item Difficult Proxy training (DP): A factor that degrades the performance of the proxy model. We calculate the error rate of proxy models using a half-size dataset.
    \item Complex Query access (CQ): Attackers find it difficult to gain query access. We halve the AP of the CA node "Query Model Access."
    \item Query Restriction (QR): If the system notices many queries greater than the minimum query calculated by the AT4EA, it blocks the account. We set the AP to zeros when the AEM node query is greater than it.
\end{itemize}

We calculate the AP by AT4EA for a system with these mitigations.
We re-compute the error rates of the classifiers with the mitigation AT because it changes the classifiers.
In addition, the DP mitigation affects the proxy model; thus, we calculate the error rates with the proxy AEA node again.
The CQ and QR require no error rate changes, but we need to calculate the APs against the difference.
We evaluate the effectiveness of these mitigations by comparing the APs of each mitigation.

All experiment codes, including those for the adversarial example matrix, EAS, and AT4EA, are publicly available on GitHub\cite{AT4EAGitHub}.

\subsection{Results}
We applid our method to the evasion attacks on different three systems to assess the versatility of itself.
We show the parameters of our method and AT4EA systematically constructed for each attack.
\begin{table}[hbtp]
    \centering
    \caption{Parameters of our method for three systems}
    \label{tab:AT4EAParam}
    \begin{tabular}{|l|c|c|c|} \hline
        Parameters & Road sign & Pneumonia & Item \\ \hline\hline
        AEMatrix & 4 & \multicolumn{2}{|c|}{11} \\ \hline
        Scenarios & 7 & 8 & 5 \\ \hline
        Depth of AT & \multicolumn{3}{|c|}{7} \\ \hline
        Nodes & 68 & 77 & 58 \\ \hline
        AEM nodes & 9 & 14 & 14 \\ \hline
        CA nodes & 23 & 28 & 18 \\ \hline\hline
        AP & $5.39 \times 10^{-5}$ & $8.58 \times 10^{-6}$ & $1.05 \times 10^{-3}$ \\ \hline
        Minimum query & 215 & 10 & 50 \\ \hline
    \end{tabular}
\end{table}
Table \ref{tab:AT4EAParam} shows the parameters of the three systems.
The experiment of road sign recognition uses the adversarial example matrix with four unique methods.
Other systems reuse the same matrix with 11 methods organized by the ART library.
Each attack scenario has all methods in its matrix.
Then, our method constructed AT4EA from each scenario using the pattern.
The table displays the total number of tree nodes. They are related to the number of their scenario nodes.
The AEA and CA nodes depend on how their scenario attacks.
We calculate the AP and minimum query for the attack by the AT4EA.
They are shown in Table \ref{tab:AT4EAParam}.
The APs indicate the vulnerability of these systems.
Furthermore, the minimum query shows that the proxy model scenario using a public dataset is the easiest black-box attack in this case.
We describe in detail of road sign recognition result in table \ref{tab:AT4EAParam}.
At first, we identify 7 EAS which include the third row of the table \ref{tab:EAS}, by using the small adversarial example matrix which has 4 methods.
Then, our method constructed AT4EA from each scenario.
A subtree in Fig \ref{fig:NodesExample} is part of it.
This AT4EA consists of 68 nodes.
The leaf AEM nodes are 9, and CA nodes are 23.
Next, we calculated the error rate of each AEM node and set the parameters each node has.
Finally, we computed the attack probability $AP=5.39 \times 10^{-5}$ and minimum query for the attack $Minimum \ query = 215$ by AT4EA.

We show the APs of the item classification system with each mitigation in the table \ref{tab:AT4EATradeOff}.
\begin{table}[hbtp]
    \centering
    \caption{Mitigation trade-off by APs}
    \label{tab:AT4EATradeOff}
    \begin{tabular}{|l|c|} \hline
        System & Attack prob.\\ \hline\hline
        Plain item classification system & $1.05 \times 10^{-3}$ \\ \hline\hline
        Adversarial Training (AT) & $1.58 \times 10^{-3}$ \\ \hline
        Difficult Proxy training (DP) & $1.04 \times 10^{-3}$ \\ \hline
        Complex Query access (CQ) & $5.38 \times 10^{-4}$ \\ \hline
        Query Restriction (QR) & $2.79 \times 10^{-5}$ \\ \hline
        AT + DP & $1.58 \times 10^{-3}$ \\ \hline
        AT + CQ & $7.95 \times 10^{-4}$ \\ \hline
        \textbf{AT + QR} & \boldmath $1.31 \times 10^{-5}$ \\ \hline
        DP + CQ & $5.38 \times 10^{-4}$ \\ \hline
        DP + QR & $2.77 \times 10^{-5}$ \\ \hline
        CQ + QR & $2.77 \times 10^{-5}$ \\ \hline
    \end{tabular}
\end{table}
The probability of the plain system became $1.05 \times 10^{-3}$ from the experiment for versatility.
We measure the effectiveness of each mitigation by comparing its probability and plain one.
The probability of the system with AT increased.
The other system probabilities decreased as below.
DP probability slightly decreased.
CQ probability became approximately half of the plain system.
QR probability became approximately $1/37$ of the plain system.
Furthermore, the probabilities of combinations of these mitigations are shown in Table \ref{tab:AT4EATradeOff}.
This suggests that the combination of AT and QR is the best mitigation from the viewpoint of the AP.

\section{Discussion}
In this experiment, we constructed AT4EA and analyzed evasion attacks against three image classification systems.
We evaluate the versatility of our method based on the experimental results.
AT4EA modeled the evasion attack on three image classification systems in the experiment.
The systems include autonomous driving, medical, and e-commerce domains.
Our method could analyze attacks on these systems.
This result suggests that our approach exhibits versatility to evasion attacks on image classification systems.
Meanwhile, there have recently been evasion attacks in other fields such as speech and malware recognition.
We will explore the applicability of our method to such fields in future work.

In the experiment of the mitigation trade-off, we calculated the APs of the systems by AT4EA.
We evaluate the effectiveness of our method based on the results.
We quantitatively analyzed the effectiveness of the mitigations using their APs calculated by AT4EA.
The system with the mitigation AT increased the AP in this case.
Applying AT allowed the classifier to become robust against proxy attacks.
The error rates of most of the methods for proxy attacks became zero using AT.
Furthermore, it prevented white-box attacks.
However, one of the methods for query attacks, which is the easiest scenario, increased its error rate in this case.
Thus, the total AP increased.
This suggests that AT is ineffective in this system.
Furthermore, the results showed that we could not expect the mitigation DP to decrease the AP.
Conversely, the mitigation CQ halved the AP.
Many scenarios have the CA "Query Model Access."
Preventing it acted as an effective mitigation to prevent evasion attacks.
In addition, the results showed that the mitigation QR was the most effective.
The minimum query for evasion attacks calculated by AT4EA determined the QR parameter.
The scenario using the query is an easy evasion attack in this system.
Therefore, QR significantly decreased the AP.
Furthermore, the minimum query for the evasion attack identified the QR parameter.
The EAS in the black-box setting requires at least 50 queries in this system.
The attacker uses the query to clarify the classifier output space to construct a proxy model using public datasets.
Conversely, all methods using the query required more than 50 queries.
For example, the SimBA method had an AP of 0.56 with 100 queries.
Therefore, we identified the QR mitigation to prevent queries greater than the minimum query.
Furthermore, the combination of these mitigations showed improved effectiveness.
The most effective combination was AT and QR.
It decreased the AP to approximately 1/80.
QR prevented query attacks.
AT made the classifier robust to white-box and proxy attacks.
Thus, their combination significantly decreased the system vulnerability.
From the above, the experiment indicates that AT4EA is effective in evaluating attack mitigations and considering their trade-offs.

\section{Conclusion}
In this study, we proposed a method for analyzing evasion attacks using attack trees.
Our method consists of the extension of conventional attack trees and the systematic construction of the extension.
The proposed extension of conventional attack trees was termed AT4EA.
It handles evasion attacks by classifying them into AEM and CA nodes.
The AEM node assigns the error rate calculated by the experiment.
The CA node assigns the AP calculated by domain knowledge.
Furthermore, we introduced nodes such as AEA and scenario nodes to construct the tree systematically.
Our AT4EA calculates the AP and minimum query for evasion attacks.
We proposed a systematic procedure to construct AT4EA.
The adversarial example matrix organizes information about AEMs from the literature.
Next, analysts identify the EAS that contains all methods in the matrix.
Our pattern translates the scenarios to AT4EA.
Finally, analysts construct CA trees under the CA nodes of AT4EA.
We can construct attack trees for evasion attacks systematically.
We conducted experiments to evaluate the versatility and effectiveness of our method.
Our approach is generally applicable to image classification systems.
Furthermore, AT4EA is effective in evaluating mitigations and considering their trade-offs.
Future works consist of research into the applicability of our method against evasion attacks in other fields and its extension to analyze other adversarial attacks.

\section*{Acknowledgement}
This work was supported by JST, CREST Grant Number JPMJCR23M1, Japan.

%--------------------------
\bibliographystyle{IEEEtran}
\bibliography{ref}

\end{document}